\documentstyle[amssym,epsfig]{mn}

\title{The primordial baryonic clouds and their contribution to the
CMB anisotropy and polarization formation.}

\author[Naselsky \& Novikov] {Pavel Naselsky$^{1,2}$ and Igor
Novikov$^{1,3,4,5}$\\ 
$^1$ Theoretical Astrophysics Center, Juliane Maries Vej 30,
DK-2100,  Copenhagen, Denmark \\ 
$^2$ Rostov State University, Zorge 5, 344090 Rostov-Don, Russia \\
$^3$ University Observatory, Juliane Maries Vej 30, DK-2100,
Copenhagen \\
$^4$ Astro-Space Center of Lebedev Physical Institute,
Profsoyuznaya 84/32, Moscow, Russia \\
$^5$  NORDITA, Blegdamsvej 17, DK-2100, Copenhagen, Denmark.}

\date{Accepted 2001 ; Received 2001 }

\begin{document}
\maketitle

\begin{abstract}

We discuss possible distortions of the ionization history of the Universe in the
model with small scale baryonic clouds. The corresponding scales of the clouds 
are much smaller than the typical galactic mass scales.  These clouds are considered in
a  framework of the cosmological model with the isocurvature and adiabatic perturbations.
In this model the baryonic clouds do not influence on the  cosmic microwave background
anisotropy formation directly as an additional sources of perturbations, but due to
change of the kinetics of the hydrogen recombination .
We also study the corresponding distortions of the anisotropy and polarization
power spectra in connection with the launched MAP and future PLANCK missions.
\end{abstract}


\newcommand{\nc}{\newcommand}
\newcommand{\beq}{\begin{equation}}
\newcommand{\eeq}{\end{equation}}
\newcommand{\be}{\begin{eqnarray}}
\newcommand{\ee}{\end{eqnarray}}
\newcommand{\num}{\nu_\mu}
\newcommand{\nue}{\nu_e}
\newcommand{\nut}{\nu_\tau}
\newcommand{\nus}{\nu_s}
\newcommand{\mnus}{M_s}
\newcommand{\taus}{\tau_{\nu_s}}
\newcommand{\nnt}{n_{\nu_\tau}}
\newcommand{\rnt}{\rho_{\nu_\tau}}
\newcommand{\mnt}{m_{\nu_\tau}}
\newcommand{\tnt}{\tau_{\nu_\tau}}
\newcommand{\bi}{\bibitem}
\newcommand{\rar}{\rightarrow}
\newcommand{\lar}{\leftarrow}
\newcommand{\lrar}{\leftrightarrow}
\newcommand{\dm}{\delta m^2}
\newcommand{\mpl}{m_{Pl}}
\newcommand{\mbh}{M_{BH}}
\newcommand{\nbh}{n_{BH}}
%\makeatletter
\def\kms{\ifmmode{{\rm km}\,{\rm s}^{-1}}\else{km\,s$^-1$}\fi}
\def\Mpc{{\rm Mpc}}

\newcommand{\eq}{{\rm eq}}
\newcommand{\tot}{{\rm tot}}
\newcommand{\M}{{\rm M}}
\newcommand{\coll}{{\rm coll}}
\newcommand{\ann}{{\rm ann}}

\section{Introduction.}

One of the most important problems of the modern cosmology is the 
determination of the density and spatial distribution of the
baryonic fraction of the matter. 

There are several sources of information about $\Omega_b h^2 =
\rho_{b}/\rho_{cr}$ parameter, where $\rho_{b}$ and $\rho_{cr}$ are
the present values of the baryonic and critical densities and $h$ is
the Hubble constant normalized to 100 $\kms\Mpc^{-1}$. Firstly, the baryonic
fraction of the matter manifests itself in the well known mass
-luminosity relation for galaxies and cluster of galaxies which leads
to the following value of the $\Omega_b h^2$ parameter: $\Omega_b
h^2\simeq 0.028^{+0.009}_{-0.008} $  (see for the review by Freedman et
al. 2001). Another one comes from the confrontation of the Standard Big Bang
 Nucleosynthesis (SBBN) theory and observational data (see for the review by
Fukugita, Hogan \& Peebles 1998). The corresponding value of the
baryonic density from this method is $\Omega_b  h^2=0.019 \pm 0.001.$
An additional empirical relation between baryonic and dark matter fractions 
$F_{b,m}=\Omega_b/\Omega_m\simeq 0.1$ at $h=0.65$  comes from X-ray
data on clusters of galaxies (Carlberg et al. 1996; Ettori \& Fabian
1999). For the most popular  $\Lambda$CDM cosmological  model with
$\Omega_m\simeq 0.3$ and $\Omega_{\Lambda} \simeq 0.7$, the corresponding
value of the $\Omega_b h^2$ parameter is $\sim 0.02$ in agreement with
the SBBN predictions.

  An independent important information about the baryonic fraction of the
matter in the Universe comes from the recent CMB experiments such as 
BOOMERANG (de Bernardis et al. 2000) and MAXIMA-1 (Hanany et al. 2000). 
Fitting the CMB anisotropy power spectrum to the above mentioned observational 
data (Tegmark \& Zaldarriaga 2000; White et al. 2000 and Lesgourgues \&
Peloso 2000) indicates that a baryon fraction parameter should be
significantly larger than the SBBN expected value, namely,  $\Omega_b
h^2\simeq 0.03$. However, recently Bond \& Critteden (2001)
show that new BOOMERANG, MAXIMA-1 and DASI data do not contradict to 
$\Omega_b  h^2=0.022 \pm 0.004$.

 It is worth noting that the above mentioned methods of the baryonic
fraction density estimation  from the CMB and  SBBN  predictions  are
based on the simple idea that the distribution of matter (including  dark
matter particles and baryons) is practically homogeneous for all
scales, except some fluctuations leading to the galaxy and large-scale
structure formation. Typically, they are assumed to be adiabatic one.  One can ask,
how sensitive are the CMB data themselves to the presence of the small--
scale baryonic (non-linear) clouds before cosmological recombination
and how can they transform the standard schemes of the cosmological
parameter extraction from the CMB data? Definitely, this possibility
is related to the isocurvature perturbations of the composite fluid
which contains baryons, CDM particles, photons and neutrinos at very
high redshift $z\gg 10^3$. There are a lot of modes of perturbations
in the composite fluid, which is discussed by Riazuelo \& Langlos (2000),
Bartolo, Matarrese \& Rioto (2001), Polarski \& Starobinsky (1994),
Abramo \& Finelli (2001), Bucher, Moodley \& Turok (2000) and others. The general
idea about classification of modes of perturbations is based on a very
simple definition of the isocurvature modes. They do not perturb the
gravitational potential. This means that the fluctuations of the total
matter density $\rho_{tot}$ are zero (see Burns 2001),
\begin{equation}
\delta\rho_{tot}=\sum\limits_{i=0}^{N}\rho_{i}\delta_i + 4\rho_{\gamma}(1+R_{\nu \gamma})
\frac{\delta T}{T}=0,
\label{eq:eq1}
\end{equation} 
where $\rho_{i}$ denotes the density of each massive species including
baryons and different kinds of the CDM particles, 
$\delta_i = \delta\rho_{i}/\rho_{i}$ is the density contrast
for each massive component, $R_{\nu \gamma}$ is the density ratio
between neutrinos $\rho_{\nu}$ and black body radiation
$\rho_{\gamma}$, and $\delta T/T$  is the CMB temperature
perturbations. 

We would like to point out that in the definition of the isocurvature
perturbations in Eq.~(\ref{eq:eq1}) one can find some peculiar mode
(or modes) which compensates the baryonic perturbations
potential, i.e., it corresponds to the condition $\rho_b\delta_b=-
\rho_x\delta_x$ for some $x$ component of the CDM particles mixture. We call below
this mode as a compensate isocurvature mode (CIM) for the
$x$-component of the dark matter particles. If several components of
the dark matter particles take part in the CIM formation, we will
continue to call them as $x$- component.\footnote{ As one can see this
mode corresponds to $\delta T/T=0$. This means that the CIM
are equivalent to an isotemperature perturbations. Note, that exactly
the same mode was described by Abramo \& Finelli (2001), but for
compensation between quintessence scalar field perturbations and some
kind of the CDM-particles.}
 
 In principle, for the CIM perturbations it is possible to assume
that the amplitudes $\delta_x$ and  $\delta_b$ are less than unity or
$|\delta_x|\sim 1$ and $|\delta_b|\simeq - (\rho_x/\rho_b)|\delta_x|\gg 1$.
One of the most interesting cases corresponds to the model with
$\delta_x\simeq -1 $, which means that some patches of the cosmological
matter do not contain the CDM $x$-particles, but at the same
patches there are non-linear clouds of the baryonic matter which compensate
the perturbations of the gravitational potential. Below we will assume that
a typical mass scale of the CIM perturbations is smaller than the
typical galactic mass scale. This means that CIM perturbations do not
influence on the CMB anisotropy formation as the additional sources of
perturbations, but they can transform the kinetics of the hydrogen
recombination. This leads to the transformation of the corresponding
$C_l$ power spectrum of the CMB  for the adiabatic fluctuations at
the scales above a few Mpc. 

 It is necessary to note that the idea about non-homogeneous distribution
of the baryonic matter at small scales is not new. The importance of
the entropic perturbations in the history of the cosmological
expansion was {\it ad hoc} demonstrated by Doroshkevich, Zel'dovich
and Novikov (1967) and Peebles (1967,1994) and recently was
generalized taking into account multi-species structure of the
cosmological plasma by Gnedin \& Ostriker (1992), Hogan (1993) \& Loeb
(1993), Peebles \& Juszkiewich (1998). The possible inhomogeneities of
the baryon fraction distribution in the epoch of the nucleosynthesis 
(Inhomogeneous Big Bang Nucleosynthesis -IBBS) was widely discussed in
the literature (see for the review by Jedamzik \& Rehm 2001) in connection
with quark-hadron phase transition. But the typical scales of such
kind of peculiarities are extremely small compared to the typical mass
scale $ M\sim  10^5 - 10^6  M_{\odot}$ for the isocurvature perturbations. 
Other events or processes have been suggested as possible sources of
the isocurvature perturbations partly connected with the baryon re-distribution
in the space. For example, cosmic strings and corresponding  currents
and magnetic fields could generate specific features in the baryonic
matter (Malaney \& Butler 1989). Yokoyama \& Sato (1991), Dolgov \&
Silk (1993), Polarsky \& Starobinsky (1994), Novikov, Schmalzing \&
Mukhanov (2000) have shown a few different ways for the generation of
the isocurvature perturbation  in the framework of the inflation
theory. 

 In connection with the above--mentioned  problem of the baryonic
fraction determination from the cosmological nucleosynthesis and the
CMB anisotropy data we dedicate our paper to the re-examination of the
models with non-linear sub-horizon-scale (at the epoch of the hydrogen
recombination) baryonic clouds with $\delta_b\gg 1$. Such
inhomogeneities do not manifest in the CMB anisotropy( because of
the extremely small scales) but manifest themselves by the
transformation of  the ionization history of the primeval
hydrogen-helium plasma at redshift $z\sim 10^3$. All these factors
should be taken into account in the reconstruction of the  ionization
history of the Universe especially at the period of the cosmological
hydrogen recombination . The reason for importance of the possible
very small scale entropy perturbations at the epoch of recombination
is connected with the very small mean free path of the Ly-$\alpha$
photons at $z\sim 10^3$: $l_{L\alpha} \sim 3\times 10^{10}((1+z)/1000)^{-5/2}
(\Omega_b h^2/0.02)^{-1}$ cm which corresponds to the baryon mass
$M_{L\alpha}=4\pi/3\rho_b l_{L\alpha}^3 \sim 2\times10^{-23}((1+z)/1000)^{-9/2}
(\Omega_b h^2/0.02)^{-2}M_{\odot}$ where $M_{\odot}$ is the Solar
mass.  High amplitude baryonic clouds with masses $M\gg M_{L\alpha}$
could transform the process of recombination at the beginning and
dissipate during recombination up to the crucial masses $M_{diss}\sim
10^5 M_{\odot}$ (Liu et al. 2001). We will show that if the typical masses of the clouds $M$ are $M>M_{diss}$ the hydrogen and helium recombination
inside and outside clouds goes independently. Due to non-linear dependency
of the electronic ionization fraction $x_e$ on the baryon density the hydrogen
and helium inside the clouds recombine  faster than  outside them.
Thus the dynamics of the mean value of the electronic ionization
fraction $x_e$ which plays a crucial role in the CMB anisotropy and
polarization formation decreases  slower than, for example, in the
uniform
model with the mean value of the baryonic fraction of the matter. We
will show that in the cloudy baryonic plasma the kinetics of the $H-He^4$
recombination is closer to the delayed recombination model by Peebles
et al. (2000) with concrete relation between $\epsilon_{\alpha}$ and
$\epsilon_{i}$ parameters of their model and
amplitudes of perturbations and the filling factor of the baryonic
clouds. We will show how sensitive the $C_l$ power spectrum of the CMB
anisotropy is to the mentioned above parameters of the
baryonic clouds. We will discuss possible manifestation of the small
scale perturbations in the MAP and upcoming PLANCK observational data.

\section{Basic definitions and modifications of the hydrogen-helium
ionization history.}

We consider a model with non-linear baryonic perturbations at the
small scales ($M\ll 10^{10} M_{\odot}$). In the analysis of the kinetics
of recombination we will take into account electrons, protons, ionized
and neutral hydrogen and helium. For simplicity we suppose that all
baryonic clouds  have the same characteristic sizes $R_{cl}$, which
are much smaller than the size of the horizon $R_{rec}$ close to the
period of recombination ($z\sim 10^3$), $ R_{cl}\ll R_{rec}$. We
denote $ \rho_{b,in}$, $\rho_{b,out}$ and $\rho_{b,mean}$ the baryon
density inside the clouds, outside of them and the mean density at the
scales much greater than $R_{cl}$ and distances between them,
respectively. We have the following relations
\begin{equation}
\rho_{b,mean}=\rho_{b,in} f + \rho_{b,out}(1-f),
\end{equation}
where $f$ is the volume fraction of the clouds. We denote
\begin{equation}
\xi= \frac{\rho_{b,in}}{\rho_{b,out}}.
\end{equation}

We can write down the following relations between mean value of the
baryon density and inner and outer values
\begin{equation}
%\begin{array}{2}
\rho_{b,in}=\frac{\xi \rho_{b,mean}}{1+f(\xi -1)},
\end{equation}
and 
\begin{equation}
\rho_{b,out}=\frac{ \rho_{b,mean}}{1+f(\xi -1)}.
%\end{array}
\end{equation}

As mentioned in Introduction the presence of the baryonic clouds in
the
primordial hydrogen-helium plasma at redshift $z\sim10^3$ changes 
the dynamics of the recombination due to non-linear dependence $x_e$ on
the baryon density. Below we will describe the kinetics of the
recombination in a cloudy baryonic fraction of the Universe taking
into account that the diffusion damping is not important
for the clouds
with $M>M_j$.
As it was shown by Liu et al. (2001), during the period of
recombination diffusion of baryons from inner to outer regions of the
clouds can suppress any small scale irregularities inside the
clumps. The natural length of this process is close to the Jeans
length  $R_J \sim c_s\eta_{rec}$ where $c_s$ is the baryonic speed of
sound and $\eta_{rec}$ is the corresponding time when the plasma became
transparent for the CMB radiation. Note that our aim is  to predict
some possible observational features in the CMB anisotropy and
polarization power spectrum connected with possible non-uniform
distribution of baryons on very small scales: much smaller than
$R_{rec}$, but greater than the Jeans scale $R_J$.
 That means that for all adiabatic perturbations at the scales 
$M\gg  10^{13}M_{\odot}$ ( which are the source of the Doppler peaks
in the CMB anisotropy and polarization power spectra ) the evolution
during the period of recombination depends not on the ionization
fraction inside or outside the clouds  but rather on the mean value of
ionization fraction over the scales of adiabatic perturbations. As we
mentioned above this mean ionization fraction does not correspond to
the ionization fraction for the mean value of the baryonic density due
to non-linear effects.  For our analysis we use two basic
software packages, RECFAST (Seager et al. 2000) and CMBFAST (Seljak \&
Zaldarriaga 1996) with modification for the cloudy baryonic model. Let us
start from CMBFAST modification because the calculation of the CMB anisotropy
and polarization  power spectra depends on the number density of free
electrons.

For the cloudy baryonic model we introduce the mean value of the
electron density 
\begin{equation}
\langle n_e \rangle= n_{e,in}f + n_{e,out}(1-f),
\end{equation}
where $f $ is the fraction of volume with clouds from Eq.(2). For the baryonic
clouds with scales $R> R_J$ we can neglect diffusion of baryons and
$Ly-\alpha$ photons and describe the recombination process inside and
outside the clouds separately. In such a case we can write down   
\begin{eqnarray}
n_{e,in} & =& x_{e,in} \left( 1-\frac{Y_{He,in}}{2} \right) n_{b,in};\nonumber  \\
n_{e,out}& =& x_{e,out} \left( 1-\frac{Y_{He,out}}{2} \right) n_{b,out},
\end{eqnarray}
where $x_{e,in}$, $x_{e,out}$, $Y_{He,in}$ and $Y_{He,out}$ are the
ionization fraction and helium mass fractions for inner and outer
regions. Note that by definition $x_e=n_e/n_H$, where $n_H$
is the number  density of neutral and ionized hydrogen. 
Let us introduce the mean value  of the ionization fraction $\langle
x_e \rangle$,
\begin{equation}
\langle x_e \rangle=\frac{\langle n_e \rangle}{\langle n_b \rangle
}\left(1-\frac{\langle Y_{He} \rangle}{2}\right)^{-1},
\end{equation}
then
\begin{equation}
\langle x_e \rangle =x_{e,in}G_{in}+x_{e,out}G_{out},
\end{equation}
where
\begin{eqnarray}
G_{in}= \frac{\xi f}{1+f(\xi -1)}
\left(\frac{1-Y_{He,in}/2}{1-\langle Y_{He} \rangle /2} \right);\nonumber  \\
G_{out} = \frac{1- f}{1+f(\xi -1)}
\left(\frac{1-Y_{He,out}/2}{1-\langle Y_{He} \rangle /2} \right),
\end{eqnarray}
and $\langle Y_{He}\rangle$ denotes the mean mass fraction of helium.

The second remark is related with the modification of the CMBFAST
code, particularly with the characteristic time of friction $\tau_D$
between electron and radiation fluids before and at the period of the 
recombination. Let us consider  the hydrodynamic equations for
baryon-electron fluid and radiation using the Newtonian approximation
. Following  Liu et
al. (2001), we have 
\begin{eqnarray}
\lefteqn{\rho^{\cdot}_b + 3H\rho_b +\nabla_{i}(\rho_bV_{i,b}) = 0,}
\nonumber  \\
&&  V^{\cdot}_{i,b} + HV_{i,b} +V_{j,b}(\nabla_{j}V_{i,b}) +
     \nabla_{i}P_b/\rho_b +    \nabla_{i}\Psi  \nonumber  \\
&=& \frac{4\rho_{\gamma} a n_e\sigma_T}{3\rho_b }(V_{i,\gamma}-V_{i,b}),
\label{eq:eqhyd}
\end{eqnarray}
where $H=a^{\cdot}/a$, $a$ is the scale factor of the Universe, $\Psi$
is the gravitational potential, dot denotes the derivative with respect to
time $\eta$  ($d\eta= dt/a$ ,  the speed of light $c=1$),
$\rho_{\gamma}$ is the density of the CMB, $ n_e$ is the local
concentration of electrons, $V_{i,\gamma}$ is hydrodynamic velocity
of radiation ( dipole moment) and $\sigma_T$ is the Thompson
cross-section. In our cloudy baryonic  model we can define
the large scale adiabatic tail of the perturbations of the matter and
the
small scale CIM tail as follows,
\begin{eqnarray}
\rho_b & = & \rho_{b,s}(\vec{r},t)+ \rho_{b,l}(\vec{r},t), \nonumber  \\
 V_{i,b} &=& V_{i,b}^{(s)}(\vec{r},t) + V_{i,b}^{(l)}(\vec{r},t), \nonumber  \\ 
\Psi & =& \Psi_s(\vec{r},t) +\Psi_l(\vec{r},t),
\label{eq:eqhyd1}
\end{eqnarray}
where 
\begin{eqnarray}
\langle \rho_b \rangle &=&\langle \rho_{b,s}(\vec{r},t) \rangle+ \rho_{b,l}(\vec{r},t), \nonumber  \\ 
\langle \rho_{b,s}(\vec{r},t)\rangle & =&\rho_{b,mean}(t),  \nonumber  \\
\rho_{b,l}(\vec{r},t)&=&\rho_{b,mean}(t)\left(1+\delta(\vec{r},t)\right),
\nonumber  \\
\langle V_{i,b}\rangle &=& \langle V_{i,b}^{(s)}(\vec{r},t) \rangle + V_{i,b}^{(l)}(\vec{r},t), \nonumber  \\
\langle V_{i,b}^{(s)}(\vec{r},t)\rangle &=&Hr_i, ~\nonumber  \\
\langle \Psi_l(\vec{r},t)\rangle &=&\langle \Psi_s(t) \rangle + \delta\Psi_l(\vec{r},t),
\label{eq:eqhyd2}
\end{eqnarray}
and $\delta(\vec{r},t)$, $V_{i,b}^{(l)}(\vec{r},t)$ and
$\delta\Psi_l(\vec{r},t)$ are the linear adiabatic perturbation of the
hydrodynamic quantities. By the definition  $\langle \ldots \rangle$
means average over the scales $|\vec{r}|\gg R_{cl}$ but
$R_{cl}|\vec{k}|\ll 1$  for all $\vec{k}$ Fourier harmonics of the
adiabatic perturbations. Using  Eq.~(\ref{eq:eqhyd1}) and
(\ref{eq:eqhyd2}) we can describe the evolution of the velocity
perturbations for the adiabatic modes 
\begin{eqnarray}
\lefteqn{\frac{\partial{V^{l}_{i,b}}}{\partial{\eta}} + HV_{i,b}^{l}  +
     \delta(\nabla_{i}P_{tot}/\rho_{tot}) +    \nabla_{i}\delta\Psi_l
     } \nonumber  \\ 
&=& \big\langle \frac{4\rho_{\gamma} a n_e\sigma_T}
{3\rho_b } \big\rangle (V_{i,\gamma}^{l}-V_{i,b}^{l}),
\label{eq:eqhyd3}
\end{eqnarray}
where $P_{tot}$ and $\rho_{tot}$ are the pressure and the density of
the baryon-photon fluid and  $\delta(\nabla_{i}P_{tot}/\rho_{tot})$
means a linear part of the perturbations. As one can see from
Eq.~(\ref{eq:eqhyd3}) in the cloudy baryonic matter the characteristic
time of baryon-radiation friction 
is
\begin{equation}
\tau_{D}^{-1}=\big\langle \frac{4\rho_{\gamma} a n_e\sigma_T}{3\rho_b
}\big\rangle = \frac{4\rho_{\gamma} a \langle x^{*}_e \rangle\sigma_T}{3 m_p },
\label{eq:eqhyd4}
\end{equation}
where $m_p$ is the proton mass. This time $\tau_{D}$ depends on the local ratio
$\langle x^{*}_e \rangle=\langle n_e(\vec{r})/n_H (\vec{r})
\rangle$. The ionization fraction $\langle x^{*}_e \rangle$ could be
written down in the following form,
\begin{equation}
\langle x^{*}_e\rangle = x_{e,in}g_{in}+x_{e,out}g_{out},
\label{eq:eqhyd5}
\end{equation}
where
\begin{eqnarray}
g_{in}& = & f\left(\frac{1-Y_{He,in}/2}{1- \langle Y_{He} \rangle /2 }\right) ;\nonumber  \\
g_{out}&=& (1-f)\left(\frac{1-Y_{He,out}/2}{1-\langle Y_{He} \rangle/
2 }\right).
\label{eq:eqhyd6}
\end{eqnarray}
Thus Eq.~(\ref{eq:eqhyd5}) and Eq.~(\ref{eq:eqhyd6})  generalize the
standard homogeneous model of the CMB anisotropy and polarization
formation on the cloudy baryonic model of the Universe.
  Note the difference between Eq.~(\ref{eq:eqhyd3}) and the standard kinetic equation for
the CMB anisotropy formation. Whereas Eq.~(\ref{eq:eqhyd3}) depends on $\langle x^{*}_e \rangle$,
the standard equation depends on $\langle x_e \rangle$,(see Eq.(8) and (9)).
The corresponding modification of the CMBFAST code
leads to modification of RECFAST programme which calculates the kinetics
of recombination inside and outside the small-scale baryonic clouds as a
function of cosmological parameter $\langle \Omega_b
\rangle=\rho_{b,mean}/\rho_{cr}$,  density contrast $\xi$ and volume
fraction $f$. In Fig.1  and Fig.2 we plot  the functions $x_{e,in}$,
$x_{e,out}$, $\langle x_e \rangle$, and $\langle x^{*}_e \rangle$ for different mean values of the baryonic density
$\langle \Omega_b \rangle$ and $h=0.65$, $\xi=11$, and $f=0.1$.  As one
can see from these figures the fractions of ionization  $\langle x_e
\rangle$ and $\langle x^{*}_e \rangle$ have different shapes and
different asymptotic at low redshifts. 
 %---------------------Figure 1 ------------------

\begin{figure}
\centering
\epsfig{file=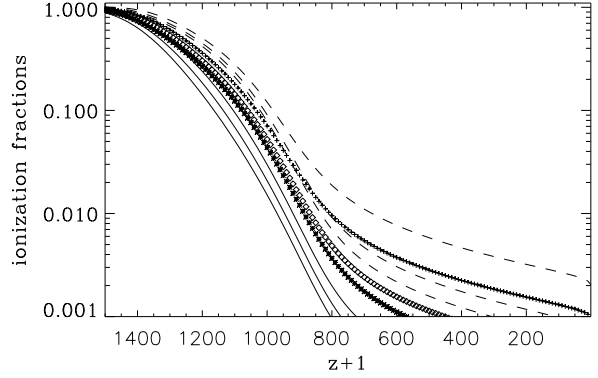,width=8.5cm}
\caption{
The fraction of ionization $\langle x_e \rangle$ as the function of $z$ for
different mean values of the baryonic density $\langle \Omega_b \rangle$ and
$\xi=11$, $f=0.1$.  From bottom to top three solid lines correspond to
the fractions of ionization for inner regions of the clouds at
$\langle \Omega_b \rangle h^2=0.03$, $\langle \Omega_b \rangle h^2 =
0.02$ and  $\langle \Omega_b\rangle h^2=0.01$
($h=0.65$). Dash lines correspond to the same but for the outer
regions and marked lines correspond to $\langle x_e \rangle$ for the same models.
 }
\label{}
\end{figure}
%---------------------Figure 1 ------------------
 
In the next section we describe the  results of the corresponding
numerical computations. However, some preliminary discussions
could be very useful for understanding of the most important
peculiarities connected with two characteristic functions of
ionization $\langle x_e \rangle$ and $\langle x^{*}_e
\rangle$. Firstly we would like to point out that in the case when the
volume fraction $f$ is small ($f\ll 1$, but $\xi f\sim1$!) the
function $\langle x^{*}_e \rangle$ should be very close to  $x_{e,out}
g_{out}$ while $\langle x_e \rangle$ practically does not change.  
%If we neglected the helium fraction of matter then 
From Eq.~(\ref{eq:eqhyd5}) and Eq.~(\ref{eq:eqhyd6})  we can 
immediately find an asymptotic of the function $\langle x_e \rangle$ at low
redshifts: $\langle x_e \rangle \simeq \langle x^{*}_{e} \rangle
G_{out}/g_{out} = \langle x^{*}_{e} \rangle /(1+f(\xi-1)) <
\langle x^{*}_{e} \rangle$. That means that the Silk
damping scale which is very sensitive on the parameter $\tau^{-1}_D$,
which is proportional to $ \langle x^{*}_{e} \rangle$, should increase due to
the increasing of the effective ionization ratio $\langle x^{*}_{e}
\rangle$. Thus we can conclude that in cloudy baryonic model the
position and amplitudes of the Doppler peaks in the $C_l$ power spectrum differ from the same
values in the standard non- cloudy model with the same mean value of
the baryon density $\langle \Omega_b \rangle$.

%---------------------Figure 4 ------------------

\begin{figure}
\centering
\epsfig{file=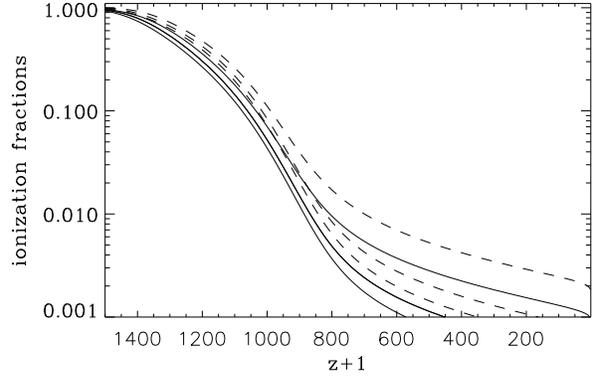,width=8.5cm}
\caption{ The fractions of ionization $\langle x_e \rangle$ and
$\langle x^{*}_e \rangle$ as the
function of $z$ for $\xi=11$, and $f=0.1$.  From bottom to top three
solid lines correspond to the fractions of ionization $\langle x_e
\rangle$ from Fig.1 for $\langle \Omega_b \rangle h^2=0.03$, $\langle
\Omega_b \rangle h^2=0.02$ and
 $\langle \Omega_b \rangle h^2=0.01$ ($h=0.65$). Dash lines from the
bottom to top correspond to ionization fractions $\langle x^{*}_e \rangle$
at the same numeration.}
\label{}
\end{figure}
%---------------------Figure 4 ------------------
Secondly, it is well known that the so-called ``visibility function'' 
$g(\tau) = \tau^{\cdot}\exp(-\tau)$, where $\tau$ is the Thompson
optical depth, depends on the function $\langle x_e \rangle$. Because
$\langle x_e \rangle < \langle x^{*}_{e} \rangle$, and $\langle
x^{*}_{e} \rangle \simeq x_{out}$ we can conclude that in our model
the kinetics of recombination is similar (in general) to the models
with additional sources of the ionization.

One additional comment is related with
the delayed recombination model by Peebles et al. (2000) mentioned in 
Introduction. We would
like to note, that the physical basis of their model and our cloudy baryonic model are completely different, but the numerical data for the
corresponding function $x_e$  in the model by Peebles et al. (2000)
and $\langle x_e \rangle$ in our model are close to each other. If we
compare the solid lines  with the marked lines in Fig.1, we can conclude that
change of the ionization fraction $\langle x_e \rangle$ plays a role in the
corresponding delay of the recombination with respect to the
recombination inside the clouds. To understand the change of the
CMB anisotropy and polarization power spectra we need to compare the
ionization fraction for outer regions and $\langle x_e \rangle$. From
Fig.1 it is clearly seen that in such a case we can introduce the term
``accelerated recombination'' because the mean ionization
$\langle x_e \rangle \rightarrow 0$  faster then $x_e \rightarrow 0$
for the outer regions at the same values of the $\Omega_b h^2$
parameter.  In both cases, following Peebles et al. (2000), we can
describe the number of additional resonance $Ly-\alpha$ quanta
produced at the epoch of the hydrogen recombination by the sources of
ionization 
\begin{equation}
\frac{dn_{res}}{dt}= \varepsilon H(t) n_b,
\label{eq:peeb}
\end{equation}
where $\varepsilon$ is the efficiency of the $Ly-\alpha$ quanta
production,  $H(t)$ is the Hubble parameter and $n_b$ is the number
density of baryons. According to Peebles et al. (2000), $\varepsilon
>0$ describes the productivity of the sources of additional resonance
quanta, which leads to the  delay of the hydrogen recombination. Formally,
if $\varepsilon < 0$ then recombination goes faster and we have
``accelerated recombination'' regime. Thus, the difference between
these two situations is related to the definition of the background
state (inner or outer parts of the clouds). Note that for the CMB
power spectrum calculations the term ``accelerated recombination'' is
preferable because it depends mainly on the characteristics of the
outer zones. 

\section{ Anisotropy and polarization power spectra  in a cloudy
baryonic Universe.}

For numerical calculations of the CMB anisotropy and polarization power spectra
we will use the modified CMBFAST code taking into account the
above-mentioned peculiarities of the ionization history of the
plasma. We take into account the difference of the $He^4$ mass
fractions $Y_{He}$ for inner and outer zones into account, using well
estimated dependence of $Y_{He}$ on ($\Omega_b h^2$)  parameter
(Olive, Steigman \& Walker 2000). For illustration of the importance
 of the cloudy structure at small scales on the
CMB anisotropy formation in Fig.3 we plot the function $\Delta^2 T
(l)= l(l+1) C_l/2\pi $ ($\mu K^2$), where $C_l$ is the
anisotropy power spectrum, for the cosmological model with
$\langle \Omega_{cdm} \rangle=0.3$, $\Omega_{\lambda} \simeq 0.65$, $h=0.65$,
$\langle \Omega_{b} \rangle h^2=0.02$ and $\Delta^2 T (l)$ for the
corresponding model with the uniform baryonic matter distribution and
scale invariant power spectrum of  the initial adiabatic
perturbations. We choose  $f=0.1$, and density contrast between inner
and outer zones $\xi=11$.

%---------------------Figure 3 ------------------

\begin{figure}
\centering
\epsfig{file=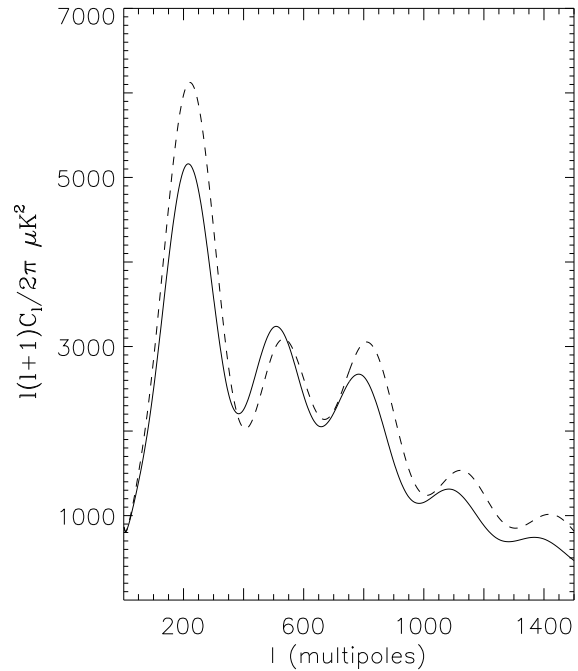,width=8.5cm}
\caption{ The $\Delta T^2 (l)$ function as a function of multipole
number $l$ for  $\xi=11$ , $f=0.1$ model.  The dash line corresponds
to the cosmological model for  $\Omega_b h^2=0.02$ without clouds.
Solid line  corresponds to the model with clouds and
$\langle \Omega_b \rangle h^2=0.02$. }
         \label{}
\end{figure}
%---------------------Figure 3 ------------------
As one can see from  Fig.3, the presence of the clouds before and
during the period of the hydrogen and helium recombination significantly
perturbed the corresponding $\Delta T^2 (l)$ function.
The same conclusion follows from Fig.4 for the polarization of the CMB
in the cloudy baryonic model.
%---------------------Figure 4 ------------------

\begin{figure}
\centering
\epsfig{file=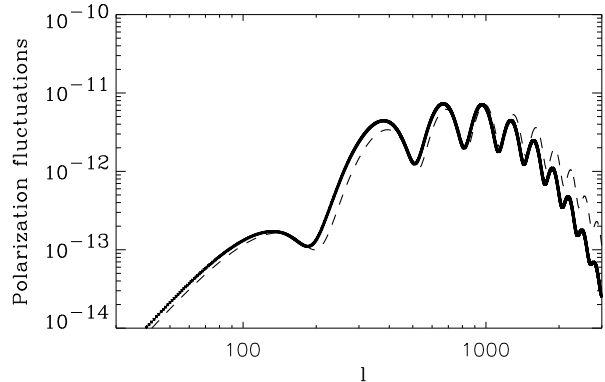,width=8.5cm}
\caption{
The polarization $\Delta^2 T_p (l)= l(l+1) C_p(l)/ 2 \pi$ as a
function  of the multipole number $l$ for  $\xi=11$, $f=0.1$ model. The
numeration of the lines is the same as in Fig.3.}

         \label{}
\end{figure}
%---------------------Figure 4 ------------------

As it is seen from Fig.4 the more complicated 
ionization history of the plasma in the cloudy model leads to the decreasing of the corresponding 
power spectrum of the polarization at $l\ge 900$ due to the increasing of the effective
dissipation scale. It is worth noting that in our cloudy model we do
not take into account the possible reionization of the hydrogen at low
redshift ($z<20$) due to the influence of the additional sources on
the ionization balance. This modification is the standard part of the
CMB anisotropy and polarization spectrum calculations using CMBFAST
codes. But it is clear that in cloudy baryonic Universe the mean value
of the optical depth must be related to the mean ionization fraction
$\langle x_e \rangle$ for corresponding values of the $\langle \Omega_b \rangle$ parameters.

\section{Conclusion.}

As it was mentioned in Introduction, the baryonic fraction of the matter is
one of the most important cosmological parameters, which determines
the most preferable model of the Universe. Two crucial parameters of
the theory are now under discussion, i.e., the light chemical
elements abundance, which is related directly to the SBBN predictions,
and the $\Omega_b$ parameter which can be determined from the current
and future CMB observations. In both cases the determination of the
parameters depends on the  hypothesis about the spatial distribution of the
baryonic fraction of the matter and can be tested by the CMB
experiments such as the launched MAP and the future PLANCK missions. 

%---------------------Figure 5 ------------------

\begin{figure}
\centering
\epsfig{file=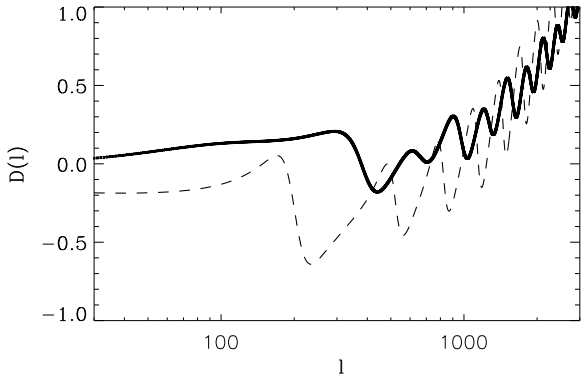,width=8.5cm}
\caption{ The functions $D(l)$ for the anisotropy (solid line) and polarization
( dashed line)  as  functions of the multipole number $l$ for
$\xi=11$ , $f=0.1$ model.  }
         \label{}
\end{figure}
%--------------------- Figure 5 ------------------

In our paper we have shown how important it is the possible   
baryonic inhomogeneity at small scales ($M \ll 10^{13} M_{\odot}$) in the
CMB anisotropy and polarization formation through distortions of the
cosmological recombination. The extremal case, when the density
contrast $\xi\simeq 10$, shows that the clumps in the baryon fraction
of the matter can  significantly change the amplitudes and positions
of the Doppler peaks in the CMB anisotropy and polarization power
spectrum. For example in the above mentioned $\xi\simeq 10$ model the
differences between $C_l$ for the anisotropy and for the polarization
in the models with $\Omega_b h^2=0.02$ ( no clouds) and $\langle
\Omega_b \rangle h^2=0.02$ (with clouds) are presented in Fig.5 in a form of the following 
the functions. For the anisotropy 
\begin{equation}
D_a(l)=2 (C_{a,nc}(l)-C_{a,c}(l))/(C_{a,nc}(l)+C_{a,c}(l)),
\end{equation}
where  $C_{a,nc}(l)$ and $C_{a,c}(l)$ denote the non-cloudy and cloudy model,
respectively, and the index $a$ corresponds to the anisotropy
spectrum. The analogous definition of the $D_p(l)$ function is used in Fig.5
for the CMB polarization. As one can see from Fig.5, practically for
all ranges of multipoles the differences between cloudy and non-cloudy
models are observable for the PLANCK mission. Moreover, if
the parameter $\xi\geq 1$, then we need to include the possible cloudy
baryonic model in the schemes of the cosmological parameter
extractions from the current and future CMB anisotropy and
polarization data. As one can see from Eq.(4) and Eq.(5), if
$f(\xi-1)\ll 1$, then the difference between $\rho_{b,out}$ and mean
baryon density is $\sim f(\xi-1) $. According to the predicted
accuracy of the cosmological parameter extraction from the
PLANCK mission, the corresponding uncertainties for the baryonic fraction
of the matter $\Delta_b= \delta\Omega_b/\Omega_b$ must be less then a
few percents. Taking conservative limit $\Delta_b\simeq f(\xi-1)\sim
0.1$ and $\xi-1 \sim 1$ we can obtain that the corresponding fraction $f$
should be detectable by the PLANCK satellite, if $f\geq 0.1$.
 
We would like to point out that our simple model of the baryonic
clouds is based on the one possible modes of the isocurvature
perturbations at the small scales. It would be interesting to
investigate the more complicated models of the initial perturbations. 
This program is in progress. 

\section*{Acknowledgments.}
Authors are grateful to A. Doroshkevich, J. Sommer-Larsen 
and J. Schmalzing for discussions. This paper is supported in part by
Danmarks Grundforskningsfond  through its support for the
establishment of the Theoretical  Astrophysics Center, by grants RFBR
17625 and INTAS 97-1192.

\end{document}